\documentstyle[preprint,prc,aps,floats,epsf]{revtex}
\begin{document}
\title{QCD at high baryon density in a random matrix model}
\author{S. Pepin}
\address{Max-Plank-Institut f\"ur Kernphysik, Postfach 103980, D-69029 
Heidelberg, Germany}
\author{A. Sch\"afer}
\address{Institut f\"ur Theoretische Physik, Universit\"at Regensburg,
         D-93040  Regensburg, Germany}
\date{\today}
\maketitle
\begin{abstract}
A high density diquark phase seems to be a generic feature of QCD.
If so it should also be reproduced by random matrix 
models. We discuss a specific one in which the random matrix elements 
of the Dirac operator are supplemented by
a finite chemical potential and by non-random elements which model the 
formation of instanton--anti-instanton molecules. Comparing our results to 
those found in a previous investigation by Vanderheyden and Jackson 
we find additional support for our
starting assumption, namely that the existence of a  high density 
diquark phase is common to all QCD-like models.
\\

\end{abstract}

\section{INTRODUCTION}
The understanding of the QCD phase diagram at finite density and temperature
is of fundamental interest to investigate phenomena related to heavy-ions
collisions and to the physics of neutron stars. While first principles lattice
calculations are able to take into account the effects of finite temperature,
the addition of a baryonic chemical potential to the lattice action
makes the fermionic determinant complex, which prevents 
reliable Monte-Carlo
calculations. Resorting to specific QCD-models or to an asymptotic QCD
expansion for very large temperature or chemical potential $\mu$ is 
thus necessary
to investigate the finite density domain. These models allow to study the
symmetries of the different regions of the phase diagram; of particular
interest is the study of the chiral restoration at finite temperature and/or
density. Recently a lot of attention has been focused on 
so-called ``colour superconductivity'': at large density and low temperature, 
an arbitrarily
weak attractive force makes the Fermi sea of quarks unstable with respect to
diquark formation and induces Cooper pairing (diquark condensation)
\cite{r1,r2,ARW98,RSSV98}.
At still higher $\mu$ this turns into still another, so-called
colour-flavour-locked, phase\cite{r5}. 
As already stated there exists no really reliable QCD  technique
to analyse these phases, except for completely unrealistically 
large $\mu$. However, all the different approaches used so far yield 
the same general picture which gives credence to the assumption that it is
generic. Some of the different approaches followed are NJL-like models 
\cite{ARW98}, which can be closely related to models based on the instanton 
phenomenology \cite{RSSV98}, the Dyson-Schwinger approach \cite{r7}, 
and, at very large $\mu$, weak coupling expansions \cite{r8}.  
These calculations agree in their main conclusions and suggest a fairly large
gap (on the order of 100 MeV), which could have some 
observational consequences in neutron stars 
and possibly even heavy-ion collisions (for recent reviews 
see \cite{Sch99,R99}). One should note however that colour-superconductivity
in QCD differs from usual superconductivity in the following sense: for 
large chemical potential, the
Renormalisation Group Approach leads to an asymptotic dependence of the 
gap as a function 
of the QCD coupling constant of the kind $\Delta \sim \exp{(-c/g)}$ 
\cite{Son99}. The naive expectation from BCS theory would be in contrast 
$\Delta \sim \exp{(-c/g^2)}$. This difference in behaviour is due to the
long-range nature of the magnetic interaction in QCD: while non-static
magnetic modes of the gluon propagator are dynamically screened due to
Landau damping, static magnetic modes are not screened. It leads to a
considerable enhancement of the QCD superconducting gap.  

To test the hypotheses that these new phases are indeed a generic feature of
QCD and QCD-like models, such that the insufficiencies of the individual 
approaches are unimportant for the existence of these phases, Random Matrix 
Theory (RMT) is the ideal tool. 
The main idea of RMT is in fact to isolate generic features.
In the past RMT was succesfully 
used to analyse the eigenvalues 
of the QCD Dirac operator as obtained in lattice gauge calculations.
For a recent review see \cite{VW00}. The perfect agreement between 
RMT predictions and lattice-QCD results led to the general conviction 
that RMT does apply to QCD,
though a rigorous formal proof is still missing.
 
RMT is especially suited to investigate the 
chiral phase transition. The Banks-Casher relation:
\begin{equation} 
<q\bar{q}> = - \frac{\pi}{V_4} \rho(\lambda) |_{\lambda = 0}
\end{equation}
relates the chiral condensate $<q\bar{q}>$ to the density $\rho$ of
zero-eigenvalues of the Dirac equation and the latter was shown to display
universal properties described by RMT.
Universal spectral correlations of the
QCD Dirac operator can also be computed at finite temperature. In
addition to giving exact results for correlations of eigenvalues, RMT can also
be used as a schematic model to investigate non-universal quantities. 

Recently Vanderheyden and Jackson have 
investigated the phase diagram 
of a QCD-like theory with generic 4-quark couplings in 
the framework of a RMT model \cite{Vdh1,Vdh2} with two quark flavours. Using 
a saddle point approximation they derived the thermodynamic 
potential for the quark and diquark condensates and
studied the competition between chiral restoration and 
diquark condensation as a function of temperature and density. They analysed
the phase diagram for different values of the coupling constants in the
diquark ($<qq>$) and chiral ($<q\bar{q}>$) channels and showed that the phase
diagram can realize a total of six general structures. 

 To model the effects of temperature, Vanderheyden and Jackson include the
lowest Matsubara frequency only. In a more recent work \cite{Vdh3}, they 
have also investigated the phase diagram of $N_c = 2$ QCD with all
Matsubara frequencies included. While the inclusion of all Matsubara 
frequencies doesn't affect the topology of the phase diagram, it eliminates
some unphysical properties like negative baryonic densities at small $\mu$
and the variation of the chiral condensate with $\mu$.
The combination of RMT and the Matsubara formalism is, however, 
only well justified if there are no relevant effects beyond the boundary 
condition in euclidian time. QCD just above the phase transition 
shows, however, strong correlations between the quarks and gluons of a
type which can hardly be described by such a simple model.
In this contribution we therefore try to improve the pioneering paper
by Vanderheyden and Jackson in this respect by allowing for a more
general type of non-random matrix elements. 
We assume that the properties of the lowest Dirac eigenstates 
are primarily determined by instanton and anti-instanton field 
configurations. 
In the instanton picture \cite{SS98}, spontaneous chiral 
symmetry breaking is assumed to be  generated by randomly distributed 
uncorrelated instantons/anti-instantons, 
which allows for a delocalization of the associated 
quark quasi-zero-modes. The restoration of chiral symmetry at high 
temperature (or high density) can then be realized if the instanton liquid 
changes from a 
random ensemble of instantons and anti-instantons to a correlated system 
where instanton and anti-instanton pair to form so-called ``molecules'' 
\cite{IS94}, the precise definition of which is ambiguous but also irrelevant 
for our purposes. The formation of such
clusters has been observed on the lattice \cite{lat99} and in numerical 
simulations of the instanton liquid \cite{ScSh96} (for more references on this
subject, see also \cite{R99}). Of course, other scenarii for chiral 
symmetry
restoration have been proposed; this phase transition can for example also
be seen as a Mott-Anderson like transition to an ``insulator state'' 
\cite{DP85}. In this work however, we assume that the dominant mechanism for
chiral symmetry restoration involves instanton molecular correlations. 
In \cite{WSW96} such a molecular model was 
used to study the chiral phase transition in the framework of RMT at zero 
density. The instanton configuration (and therefore the temperature) was
characterized by two parameters. 
It was found that to reduce the value of the chiral
condensate by more than a factor two, about 95 percent of the instantons and
anti-instantons have to pair in molecules, which confirmed the decisive role
played by molecules formation in the restoration of chiral symmetry at finite
temperature. 
In this paper we want to generalize the results 
of \cite{WSW96} to finite chemical potential. That will allow us to 
investigate the stability of the results found in \cite{Vdh1,Vdh2} when the 
temperature effects are not modelled in the most elementary manner by the
first Matsubara frequency. Within RMT the properties of the chiral phase 
transition are determined by the interplay between fluctuations (described by
the random matrix elements) and constant terms (for fixed $T$ and $\mu$) in 
the Dirac operator. In \cite{Vdh1,Vdh2} the latter were assumed to be the 
same for all states, while we allow for the possibility that they only 
contribute for a certain fraction of them. We find that for small 
$\mu$ the existence of a phase transition is extremely sensitive to 
this fraction (and thus to the detailed instanton-anti-instanton dynamics) 
while it becomes nearly independent of it for  large $\mu$,
suggesting that the transition to a diquark condensate is indeed a 
model-independent feature.

\section{FORMULATION OF THE MODEL}
Apart for the parametrization of temperature we follow closely the approach
adopted by Vanderheyden and Jackson \cite{Vdh1,Vdh2} and refer to their 
original papers for details of the derivation of the thermodynamic
potential. Our starting partition
function (for two quark flavours $\psi_1$ and $\psi_2$) is the following:
\begin{equation}
\begin{array}{l}
Z(\mu,d,\alpha) = \displaystyle \int {\cal D H} {\cal D}\psi_1^{\dag} 
{\cal D}\psi_1
{\cal D}\psi_2^{T} {\cal D}\psi_2^* \\
\times \exp{\left[ i \left(\begin{array}{c} \psi_1^{\dag} \\ \psi_2^{T}
\end{array} \right)^T \left( \begin{array}{cc} {\cal H} + (D + i \mu)
\gamma_0 + i m & \eta P_{\Delta} \\ -\eta^* P_{\Delta} &
-{\cal H}^T + (D - i \mu) \gamma_0^T - i m \end{array} \right) 
\left(\begin{array}{c} \psi_1 \\ \psi_2^{*} \end{array} \right) \right] }
\end{array}
\label{part}
\end{equation}
i.e., it is the same as in Eq. (1) of \cite{Vdh2} but with the substitution
$\pi T {\bf 1}_N \rightarrow D$ with $D$ being a $N \times N$ 
diagonal matrix, for which a
fraction $\alpha$ of its diagonal
elements are equal to a fixed value
$d$ and the rest is zero. 
This form is motivated by the observation that the correlation of instantons 
and anti-instantons (i.e. the formation of molecules) generates
such diagonal terms. The size of these terms depends on the 
anti-instanton-instanton separation and $d$ has to be interpreted as 
an average value.
If $\alpha =1$ (that is, if all instantons and anti-instantons form 
molecules), one recovers the model of Vanderheyden and Jackson.
Let us stress that there are good arguments to assume that 
for the physical phase transition $\alpha$ is in fact close to 1.
We shall discuss some of them in the conclusion.

In (\ref{part}), $m$ is the current mass of the quark and $\eta$ the 
source term for the
diquark condensate $<\psi_2^T P_{\Delta} \psi_1>$, where $P_\Delta \equiv i C 
\gamma_5 \lambda_2$ projects on a colour $\bar{3}$, scalar diquark state ($C$
is the charge conjugation matrix). 
$\eta$ has to be taken to zero at the end of the calculations. A general
hermitian interaction ${\cal H}$ can be written as an expansion over the 
sixteen Dirac matrices $\Gamma_K$ times the $N_c^2$ colour matrices 
$\lambda^a$:
\begin{equation}
{\cal H}_{\lambda i \alpha k ; \kappa j \beta l} = \sum_{K=1}^{16} 
(\Gamma_K)_{\lambda i ; \kappa j} \sum_{a=1}^{N_c^2} \lambda^a_{\alpha\beta}
(A^{K a}_{\lambda \kappa})_{kl}
\end{equation}
The measure ${\cal D H}$ associated with the random matrices $A^{K a}$ is:
\begin{equation}
{\cal D H} = \{ \prod_{K a} \prod_{\lambda \kappa} 
{\cal D}A^{K a}_{\lambda 
\kappa} \} \exp{\Bigl[- N \sum_{K a} \sum_{\lambda \kappa} \beta_K 
\Sigma_{K a}^2
Tr[A^{K a}_{\lambda\kappa} (A^{K a}_{\lambda\kappa})^T] \Bigr]}
\end{equation}  
with ${\cal D}A^{K a}_{\lambda \kappa}$ being 
the Haar measure. $\beta_K=1$ if $K$ 
corresponds to vector or axial-vector interaction and $\beta_K=1/2$ if $K$
is scalar, pseudo-scalar or tensor. The variance $\Sigma_{K a}$ is the same
for all channels.  
 
Following \cite{Vdh1,Vdh2}, one first performs the integration over the random
matrix interaction and then uses a Hubbard-Stratonovitch transformation to
introduce two auxiliary variables $\sigma$ and $\Delta$, associated 
respectively with the chiral and diquark condensate. After integrating out the
fermion fields, one obtains the following partition function:
\begin{equation}
Z(\mu,\alpha,d) = \int d\sigma d\Delta \exp{\left[-4N \Omega(\sigma,\Delta)
\right]}  
\end{equation}
with the thermodynamic potential
$\Omega(\sigma,\Delta)$:
\begin{equation}
\begin{array}{r}
\Omega(\sigma,\Delta) = A \Delta^2 + B \sigma^2 - \frac{1}{2} \left\{ \alpha
(N_c -2) \ln{\Bigl[ ((\sigma + m +\mu)^2 + d^2) ((\sigma + m -\mu)^2 + d^2)
\Bigr]}  \right. \\
+ (1 - \alpha) (N_c - 2) \ln{ \Bigl[(\sigma + m +\mu)^2 (\sigma + m -\mu)^2 
\Bigr]} \\
+ 2 \alpha \ln{\Bigl[ ((\sigma + m +\mu)^2 + d^2 + \Delta^2) ((\sigma + m 
-\mu)^2 + d^2 + \Delta^2) \Bigr]} \\ \left.
+ 2 (1 - \alpha)  \ln{\Bigl[ ((\sigma + m +\mu)^2 + \Delta^2) 
((\sigma + m -\mu)^2 + \Delta^2) \Bigr]} \right\}
\end{array}
\label{pot}
\end{equation}
As discussed in \cite{Vdh1,Vdh2}, the couplings $B$ and $A$ are weighted
averages of the Fierz coefficients obtained by projection of the original 
interaction on chiral and scalar diquark channels respectively.
The ratio $B/A$ is the only independent parameter and measures the balance 
between 
chiral and diquark condensation; varying this ratio allows to explore all the
different possible structures of the phase diagram. Imposing the interaction
to be hermitian gives the upper bound $B/A \leq N_c/2$. The absolute 
magnitudes of $A$ and $B$ play a secondary role. They introduce a scale for 
the condensating fields but don't affect the structure of the phase diagram.
In the following calculations, we fix $A=1$ and vary $B$.  

To study the various phases for a given value of $B/A$ one must minimize the 
potential $\Omega$ with respect to both condensates and solve the resulting
gap equations. One can already notice 
that, if $\alpha \neq 1$, the second term of this potential is singular at 
zero density if the chiral condensate vanishes,
$\sigma=0$ (the current quark mass $m$ will be set to
zero in the following calculations). This is due to the fact  
that chiral symmetry cannot
be completely restored at zero density unless all the instantons and 
anti-instantons form molecules (i.e. $\alpha=1$), 
see also \cite{WSW96}. 

\section{Results and discussion}
\begin{figure}
\epsfysize=12cm \centerline{\epsffile{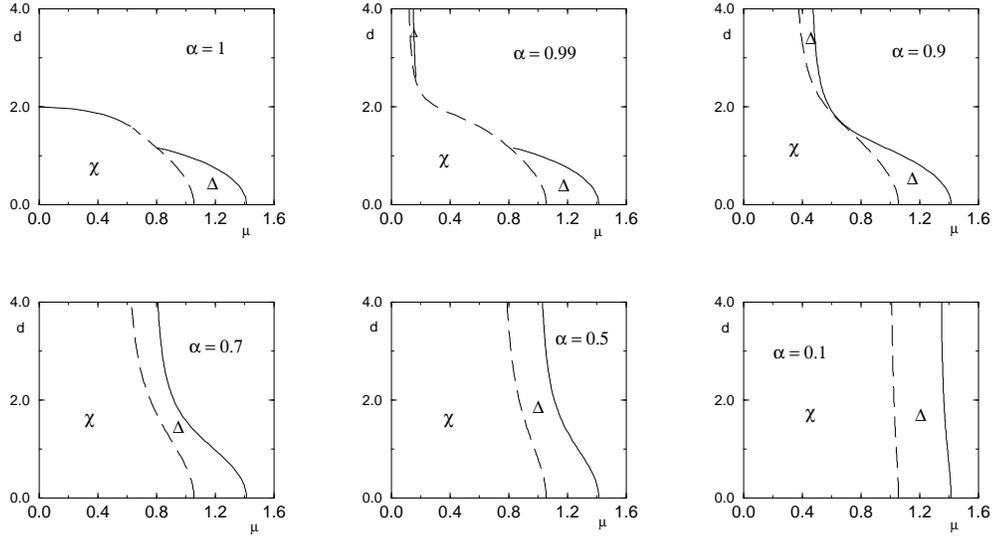}}
\caption{Phase diagram for B/A=0.75. The x-axis corresponds to the chemical
potential; the y-axis to the parameter $d$. $\chi$ and $\Delta$ label the 
chiral and diquark phase respectively. Dashed and continuous lines correspond 
respectively to first and second-order phase transitions. Note that 
QCD-phenomenology suggests that at the phase transition $\alpha$ is close to 1.
\label{phase075}}
\end{figure}

\begin{figure}
\epsfysize=12cm \centerline{\epsffile{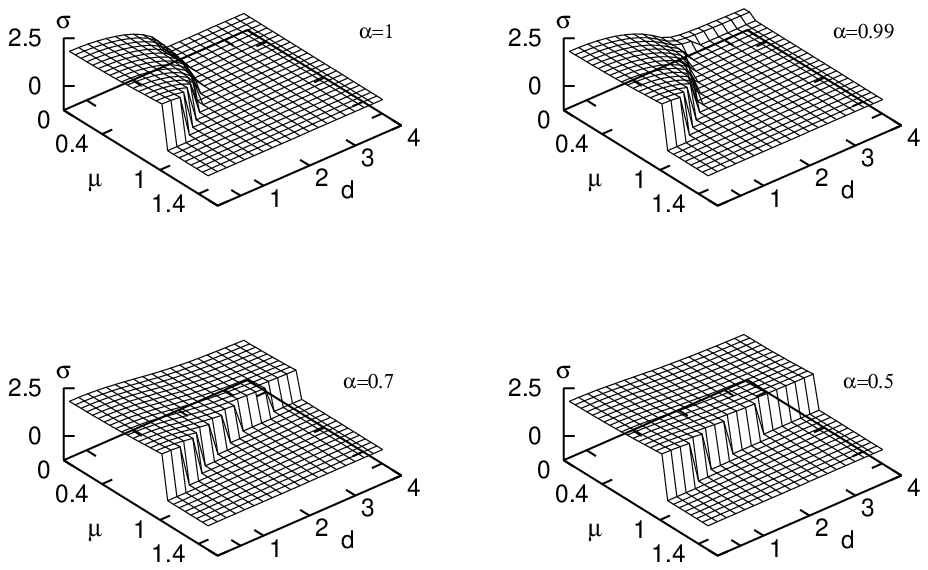}}
\caption{The strength of the quark condensates $\sigma$ in Fig.\ref{phase075}
($B/A = 0.75$) 
\label{qcon}}
\end{figure}

\begin{figure}
\epsfysize=12cm \centerline{\epsffile{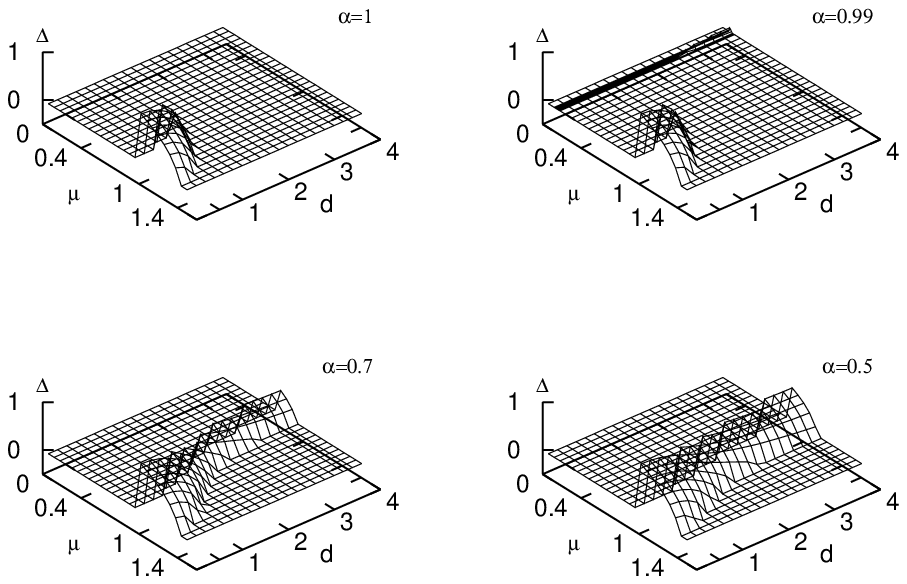}}
\caption{The strength of the diquark condensates $\Delta$ in Fig.\ref{phase075}
\label{dqcon}}
\end{figure}

The gap equations derived from the potential (\ref{pot}) admit four kinds of
solutions: the trivial vacuum where both condensates $\sigma$ and $\Delta$ 
are zero; the chirally-broken phase where $\sigma \neq 0$ 
and $\Delta=0$; the colour-superconducting phase
($\Delta \neq 0, \sigma = 0$); the mixed-phase, where
both condensates are non-zero. By varying the ratio $B/A$, we recovered the 
various scenarios discussed 
in \cite{Vdh2}; in particular, the mixed-phase appears only for 
$B/A \geq 1.05$. We 
first present results for the 
coupling ratio $B/A = \displaystyle \frac{1}{2} (\frac{N_c}{N_c-1}) = 
0.75$ 
corresponding to one-gluon-exchange as well as to 
an instanton-induced interaction. Fig.\ref{phase075} shows the
phase diagram in the $\mu - d$ plane for six different values of the 
molecule fraction $\alpha$: 1, 0.99, 0.9, 0.7, 0.5 and 0.1. 
Such phase diagrams can be misleading if the actual condensate values are 
too small to be stable against fluctuations. This does, however, not seem to 
be the case, as can be seen from Fig.\ref{qcon} and \ref{dqcon}.
  
For $\alpha=1$, one recovers the results of
\cite{Vdh2} by identifying the parameter $d$ with the first Matsubara 
frequency $\pi T$. We have checked that the restoration of chiral symmetry 
is first-order for small
enough $d$ ($d < 1.57$). The phase diagram changes noticeably for 
$\alpha \neq 1$: as anticipated
previously, chiral symmetry cannot be restored at zero-density. 
(This is well known, as $\alpha\neq 1$ implies the presence of 
isolated instantons and thus zero-modes.) 
The chiral
transition becomes first-order for all values of $d$, so there is no 
longer a  
tricritical point in the phase diagram.
Moreover the diquark phase appears
also at large $d$ and relatively low density. If $\alpha$ is 
further decreased, 
the diquark phase extends to all values of $d$ and for
growing values of $d$ the chiral 
symmetry restoration occurs at higher  
$\mu$.
One notes, however,  
that chiral symmetry restoration always occurs for  $\mu$
above a certain critical value, even
if the fraction of instanton-molecules is small. 
We conclude  that for small $\mu$ the properties of the chiral 
phase transition depend crucially on the detailed 
instanton-anti-instanton dynamics. In contrast,
the occurrence of a diquark phase seems to be a
generic property of the phase diagram whatever the fraction of molecules.
    
\begin{figure}
\epsfysize=12cm \centerline{\epsffile{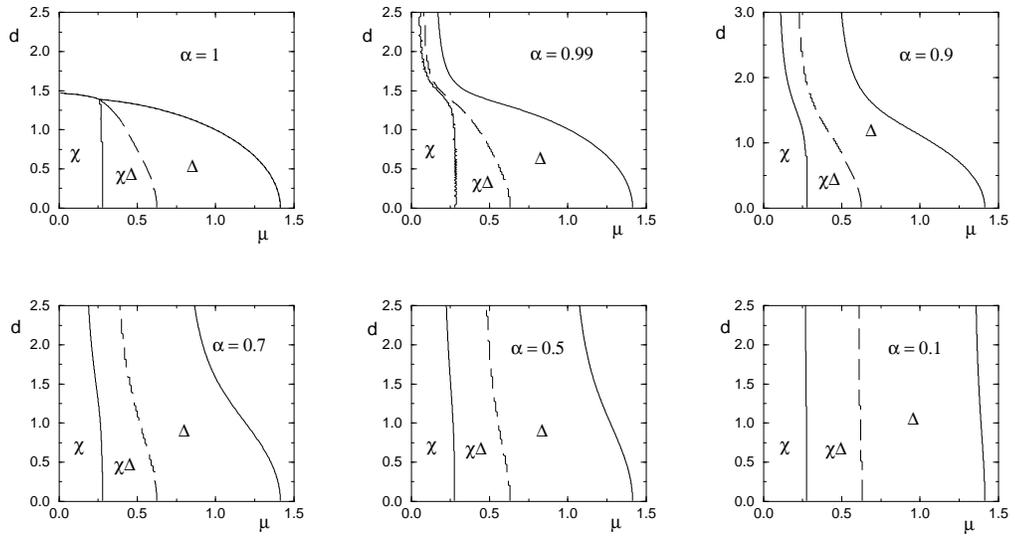}}
\caption{Same as Fig. 1 but for B/A=1.4. $\chi\Delta$ labels the mixed-phase.
\label{phase14}}
\end{figure}

For larger value of the ratio $B/A$ a mixed phase appears: we show in 
Fig.\ref{phase14} results for the case $B/A = 1.4$. 

Again we reproduce the results of 
\cite{Vdh2} in the case $\alpha = 1$. The transition from the mixed-phase 
$\chi\Delta$ to the diquark phase $\Delta$ is first-order for $d < 1.1$. 
For $\alpha < 1$, the changes 
in the phase diagram are similar to the ones observed in Fig.\ref{phase075}. 
The mixed-phase and the diquark phase extend over the whole range of $d$ 
and the
transition from the mixed-phase to the diquark phase is now first order for
all values of $d$. As in 
the previous case the diagram is stable only for high enough $\mu$. 

For completeness, we have also checked the results obtained in \cite{Vdh2} when
the current quark mass $m$ is  non-zero. For $\alpha=1$, one 
recovers their phase diagram, where a first-order transition line 
(corresponding to the first-order chiral transition of the chiral limit) ends
in a critical point. This critical point 
however disappears if $\alpha \neq 1$
and one finds a first-order transition for all $d$.
   
Before to conclude this section, we consider briefly the case $N_c=2$. 
As it has been shown in \cite{Vdh1,Vdh2}, the ratio $B/A$ is necessarily equal 
to one in that case. At zero chemical potential, the potential $\Omega$ 
depends on the condensation fields $\sigma$ and $\Delta$ through the 
combination $\sigma^2 + \Delta^2$. As soon as 
$\mu \neq 0$, this symmetry is broken and the chiral condensate vanishes; the
system prefers diquark condensation over chiral symmetry breaking.
The same conclusions are valid
for a molecule fraction $\alpha \neq 1$: the only difference is that the 
diquark condensate stays non-zero at low $\mu$ even for large values of $d$.
\section{Conclusions}
We have investigated the phase diagram of QCD as a function of density and
temperature within Random Matrix Theory along the lines of \cite{Vdh1,Vdh2}.
The aim was to distinguish generic and specific properties. We did not 
treat 
the temperature dependence explicitely but  
encoded it in the fraction $\alpha(\mu,T)$ and strength
$d(\mu,T)$ of instanton-anti-instanton correlations. 
We studied the stability of the phase diagram
with respect to variations of $\alpha$ and $d$ in comparison with the
results of \cite{Vdh2}. We found
that the low-density part of the phase diagram is highly 
dependent on $\alpha(\mu,T)$ and $d(\mu,T)$ and thus on
the detailed instanton dynamics, while,
 on the other hand, the phase 
diagram is rather stable for high
densities. 
We also found that for finite $\mu$ not all instantons and anti-instantons have
to combine into molecules to allow for a chiral phase transition.
These are our main results. Of course, Random Matrix Theory only allows us
to derive qualitative results. In particular, it can not address the 
question mentioned in
the introduction about the dependence of the gap on 
the QCD coupling constant; this dependence is closely related to the 
behaviour of the gluon propagators while our approach amounts to use a 
contact 
quark interaction.

For quantitative studies we need realistic models for
the functions $\alpha(\mu,T)$ and $d(\mu,T)$ i.e. specific input 
which goes beyond RMT. The
fraction of molecules has been calculated as a function of temperature at
zero density in \cite{VS96}, where it was shown that the fraction of 
molecules jumps rapidly from $\alpha \sim 0.5$ at $T=0.7~ T_c$ to $1$ at
$T_c$. In \cite{R99,R98} the molecule density was computed as a function of 
the chemical potential at zero temperature. It was found that there is a 
delicate competition between random instantons engaged in diquarks and 
molecule formation: in fact, the former are dominant and induce the chiral 
transition. The fraction of atomic instantons goes, however, also in this 
case to  zero  
when chiral symmetry is restored. In our case also, the diquark phase 
appears in the regions of the phase diagram where the effects of molecules 
are weak enough (meaning either a small fraction $\alpha$ or a relatively 
low value of $d$). 

 Relating our phase diagrams to the one in the ($\mu,T$) plane 
is not
a trivial question. We have shown that the phase diagram is rather unstable
at low density, which means that its characteristics in the ($\mu,T$) plane 
should 
depend on the details of the functions $\alpha(\mu,T)$ and $d(\mu,T)$. The 
results quoted above indicate that the fraction of molecules
$\alpha$ changes abruptly at the phase transition. It is natural to assume 
that this will also happen for the intermediate case ($T$ and $\mu$ non-zero),
 which suggests a large value of $\alpha$. Whether this value will be
exactly equal to one (which, as we have seen, is necessary for a vanishing 
of the chiral condensate) or not is not known.
However, if the value of $\alpha$ is large enough ($> 0.9$),
the chiral condensate will anyway not be stable against 
fluctuations. We feel that our present understanding of the QCD phase
transition is insufficient for reliable quantitative models for the
functional form of $\alpha(\mu,T)$ and $d(\mu,T)$.  
 
This work was supported by BMBF(GSI) and DFG. We acknowledge helpful 
discussions with M. Alford, A. Jackson, B. Vanderheyden, and 
H.A. Weidenm\"uller.


\begin{references}
\bibitem{r1} B. Barrois, Nucl. Phys. B129 (1977) 390.
\bibitem{r2} D. Bailin and A. Love, Phys. Rep. 107 (1984) 325, and
references therein.
\bibitem{ARW98} M. Alford, K. Rajagopal and F. Wilczek, Phys. Lett. B422 
(1998), 247.
\bibitem{RSSV98} R. Rapp, T. Sch\"afer, E.V. Shuryak and M. Velkovsky,
Phys. Rev. Lett. 81 (1998), 53.
\bibitem{r5} M. Alford, K. Rajagopal and F. Wilczek, Nucl. Phys. B537 (1999) 
443.
\bibitem{r7} D.K. Hong, V.A. Miransky, I.A. Shovkovy, and L.C.R. 
Wijewardhana, Phys. Rev. D61 (2000) 056001, Erratum D62 (2000) 059903,
and references therein.
\bibitem{r8} R.D. Pisarski and D.H. Rischke, Phys. Rev D61 (2000) 074017,
and references therein.
\bibitem{Sch99} T. Sch\"afer, nucl-th/9911017. 
\bibitem{R99} R. Rapp, T. Sch\"afer, E.V. Shuryak and M. Velkovsky, Annals
Phys. 280 (2000) 35.
\bibitem{Son99} D.T. Son, Phys. Rev. D59, (1999) 094019.
\bibitem{VW00} J.J.M. Verbaarschot and T. Wettig, hep-ph/0003017.
\bibitem{Vdh1} B. Vanderheyden and A.D. Jackson, Phys. Rev. D61 (2000) 076004.
\bibitem{Vdh2} B. Vanderheyden and A.D. Jackson, Phys. Rev. D62 (2000) 094010.
\bibitem{Vdh3} B. Vanderheyden and A.D. Jackson, hep-ph/0102064.
\bibitem{SS98} T. Sch\"afer and E.V. Shuryak, Rev. Mod. Phys. 70 (1998), 323.
\bibitem{IS94} E.-M. Ilgenfritz and E.V. Shuryak, Phys. Lett. B 325 (1994)
 263.
\bibitem{lat99} P. de Forcrand, M. Garcia Perez, J.E. Hetrick, E. Laermann,
J.F. Lagae, I.O. Stamatescu, Nucl. Phys. Proc. Suppl. 73 (1999), 578.
\bibitem{ScSh96} T. Sch\"afer and E.V. Shuryak, Phys. Rev. D53 (1996), 6522.
\bibitem{DP85} D.I. Diakonov and V.Y. Petrov, Sov. Phys. JETP 62 (1985), 204. 
\bibitem{WSW96} T. Wettig, A. Sch\"afer and H.A. Weidenm\"uller, Phys. Lett.
B367 (1996), 28. 
\bibitem{VS96} M. Velkovsky and E. Shuryak, Phys. Rev. D56 (1997), 2766.
\bibitem{R98} R. Rapp, Nucl. Phys. A642 (1998), 71.
\end{references}
\end{document}